\documentclass[12pt]{article}
\usepackage{axocolor}
\usepackage{graphicx}
\usepackage[usenames]{color}
\usepackage{bm}
\usepackage{dcolumn}
\usepackage{epsfig}

\newcommand{\ba}{\begin{array}}
\newcommand{\ea}{\end{array}}
\newcommand{\bd}{\begin{displaymath}}
\newcommand{\ed}{\end{displaymath}}
\newcommand{\beq}{\begin{equation}}
\newcommand{\eeq}{\end{equation}}
\newcommand{\bea}{\begin{eqnarray}}
\newcommand{\eea}{\end{eqnarray}}

\def\tbw{\beta_W}
\def\tbV{\beta_V}

\def\dis{\displaystyle}
\def\barr{\begin{array}}
\def\earr{\end{array}}
                               
                              \def\gev{\: \rm GeV} 
                              \def\tev{\: \rm TeV} 
                              
                              \def\fb {\: \rm fb}
\def\rbw{\mathcal R(b_W)}
\def\rbetw{\mathcal R(\beta_W)}
\def\ibw{\mathcal I(b_W)}
\def\ibetw{\mathcal I(\beta_W)}

\newcommand{\beqn}{\begin{eqnarray}}
\newcommand{\eeqn}{\end{eqnarray}}

\def\q2 {q^2}

\def\lra{\longrightarrow}

\title{Anomalous Higgs Couplings at an $e \, \gamma $ Collider}
\author{Debajyoti Choudhury \\
       {\it {\small Department of Physics and Astrophysics}},\\
             {\it {\small University of Delhi, Delhi-110007, India}}.\\
              {\small and}\\
              Mamta\\
             {\it {\small Department of Physics}},\\
                   {\it {\small S.G.T.B. Khalsa College}},\\
                    {\it {\small University of Delhi, Delhi-110007, India.}}
       }

\date{}

\begin{document}   

\maketitle
\begin{abstract}
\noindent
We examine the sensitivity of $ e \; \gamma$ colliders (based on $e^+
\; e^- $ linear colliders of c.m. energy 500 GeV) to the
anomalous couplings of the Higgs to W-boson via the process $e^- \;
\gamma  \longrightarrow \nu \; W\; H $. This has the advantage over
$ e^+ \; e^-$ collider in being able to dissociate WWH vertex from
ZZH. We are able to construct several dynamical variables which may be
used to constrain the various couplings in the WWH vertex.
\end{abstract}

\newpage
\section{Introduction}
\noindent
The Standard Model (SM) of particle physics, based on the gauged
symmetry group $ SU(3)_C \times SU(2)_L \times U(1)_Y$, has proven to
be incredibly successful in describing the electromagnetic, weak and
strong interactions. However, the precise mechanism of the electroweak
symmetry breaking and mass generation still remains one of the
important open questions of the theory. Within the SM, the breaking of
symmetry is realized \textit{via} the Higgs mechanism in which a
scalar SU(2) doublet, the Higgs boson (which is not yet an
experimental reality) is introduced \emph{ad hoc} and the symmetry
spontaneously broken by virtue of the Higgs field acquiring a vacuum
expectation value.  However, in this realization, the theory suffers
from the ``naturalness'' problem since the running Higgs mass is
quadratically divergent necessitating a fine tuning in order to keep
the theory perturbative. Conversely, this implies the existence of a
cut-off scale $\Lambda$ (widely believed to be of the order of TeV)
above which new physics must appear.

Probing the mechanism of EWSB and the search for a Higgs boson
together constitute one of the main goals of present and future
TeV-scale colliders.  The direct search for the Higgs boson in the
CERN LEP experiments sets a lower bound on its mass of $m_H>114.4 \,$
GeV \cite{higmin}. The precision electroweak data, on the other hand,
favor a light Higgs boson with a mass $m_H \leq 186 \,$ GeV at 95\%
CL \cite{higmax}. It should be noted though that both these bounds are
model-dependent and, in theories going beyond the SM, maybe modified
to a significant degree.  For example, in two-Higgs doublet models,
with~\cite{2hdm} or without supersymmetry~\cite{cp-mssm}, the lower
limit from direct searches at LEP and elsewhere is still as low as 10
GeV~\cite{lowerH}.  Similarly, the upper bound on the mass of the
(lightest) Higgs in some extensions may be substantially
higher~\cite{highH}.  The Large Hadron Collider (LHC) is expected to
be capable~\cite{lhc-tdr} of searching for the Higgs boson in the
entire mass range allowed.

In case a Higgs boson is found at TeV-scale colliders, it is of
fundamental importance to check if the Higgs boson is SM-like by
studying its couplings to the SM particles. In particular, if no new light
particles other than the Higgs boson are detected in the
next generation collider experiments, it is even more pressing to
determine the Higgs boson couplings as accurately as possible to look
for hints for new physics beyond the SM.

As the dominant neutral Higgs production modes at a linear collider
proceed via its coupling with a pair of gauge bosons ($VV, \ V=W,Z$),
these are expected to be sensitive to the $VVH$ couplings, and
departures from their SM values can be probed via such production
processes.  Kinematical distributions for the process $e^+e^-\to f
\bar f H$, proceeding via vector boson fusion and Higgsstrahlung have
been studied both without~\cite{dist,bcgs} and with beam
polarization~\cite{pol,bcgs}.  Anomalous $ZZH$ couplings, expressed in
terms of higher dimensional operators, have been studied in
Refs.\cite{rattazzi,stong,He:2002qi, Hagiwara:1993ck,Gounaris:1995mx}
for the LC and in Refs.\cite{rosenfeld,plehn,Zhang:2003it} for the
LHC.  And whereas Ref.~\cite{Hagiwara:2000tk} probes the anomalous
$ZZH$ and $\gamma Z H$ couplings using the optimal observable
technique~\cite{oot}, Refs.~\cite{bcgs,t.han}, on the other hand, use
asymmetries in kinematical distributions. In Ref.~\cite{maria}, the
$VVH$ vertex is studied in the process of $\gamma\gamma \to H\to
W^+W^-/ZZ$ using angular distributions of the decay products.

While the aforementioned studies have well established the high
resolving power of the $e^+ e^-$ linear collider in resolving the
$ZZH$ vertex, the sensitivity to the $WWH$ vertex is not as good. As
Ref.\cite{bcgs} explicitly exhibits, observables depending on the
latter most often also receive contributions from the former ($ZZH$),
thereby making it very difficult to untangle the two. Furthermore, the
leading order process at an $e^+ e^- $ collider sensitive to the $WWH$
vertex, namely the $\bar \nu_e \nu_e H$ production channel, has too
few observables associated with it. It is thus contingent upon us to
look for alternative channels with enhanced sensitivity to this
vertex.

A high energy $e\; e$ linear collider provides just such a theater in the form
of a high energy photon beam option. As the electron (positron) bunches in
these colliders are used only once, it is possible to convert electrons to
real high-energy photons using the Compton back-scattering of laser light and
thus obtain $\gamma\; \gamma$ and $e \, \gamma $ colliders with real
photons. With the luminosity and energy of such colliders being comparable to
those of the basic $e\; e$ collider \cite{telnov}, one may now consider a
process such as
\beq 
e^- + \gamma \to \nu_e + W^- + H \ .
    \label{the_process}
\eeq
Clearly, this process is sensitive to the $WWH$ vertex, but not to the
$ZZH$ one. Furthermore, with both the Higgs and the $W$ being visible
(in their decay modes), one is offered a plethora of kinematical
variables in the construction of suitable observables. Thus, this
process suffers from neither of the two aforementioned problems that
plagued the dominating channel at the parent $e^+ e^-$ machine.

The outline of the paper is as follows: In section \ref{sec:vvh}, we
discuss the possible sources and symmetries of various VVH couplings
and the rate of the process used to constrain these couplings. A
realistic experiment and the acceptance cuts are discussed in section
\ref{sec:real_expt}.  In section \ref{sec:unpol}, we construct several
observables to constrain the WWH vertex using unpolarized beams at
c.m. energies of 500 GeV. The effect of polarized beams is discussed in
section \ref{sec:pol}. In section \ref{sec:PT-conjugate}, we use the
conjugate process ($e^+ \; \gamma \lra \bar \nu \; W \; H$)
to improve the limits.

\section{VVH Couplings}
\label{sec:vvh}
Within the SM as well as its minimal supersymmetric counterpart (MSSM), the
only (renormalisable) interaction term involving the Higgs boson and a pair of
gauge bosons is the one arising from the Higgs kinetic term. However, once we
accept the SM to be only an effective low energy description of some other
theory, higher dimensional (and hence non renormalisable) terms are also
allowed.

Demanding only Lorentz invariance and gauge invariance, the most
general coupling structure may be expressed as
\beq
 \Gamma_{\mu \nu}^V = g_V \left[ a_V g_{\mu \nu} + \frac{b_V}{m_V ^2}
 \left( k_{2 \mu} k_{1\nu} - g_{\mu \nu} k_1 \cdot k_2 \right) + 
 \frac{\tbV}{ m_V ^2} \epsilon_{ \mu \nu \alpha \beta}
 k_1^\alpha k_2^\beta \right] 
\label{vertex}
\eeq
where $k_1^\mu$ and $k_2^\nu$  are the momenta of two $W$'s (or $Z$'s) with 
\bea
 g_W^{SM} &=& e\;\cot{\theta_W}\; M_Z \nonumber \\
 g_Z^{SM} &=& 2 \; e\; M_Z / \sin{2 \theta_W}.
\eea
  In the context of the SM, at the tree level, $a_W^{SM} =
a_Z^{SM}=1$, while the other couplings vanish identically.  At the
one-loop level or in a different theory, effective or otherwise, these
may assume significantly different values.

In general, each of these couplings can be complex, reflecting
possible absorptive parts of the loops, either from the SM or from
some new high scale physics beyond the SM. Note, though, that in most
processes of interest wherein the amplitude is linear and homogenous
in the $VVH$ couplings ({\em i.e.}, when the Yukawa couplings may be
neglected), an overall phase of the couplings is irrelevant. Thus, one
phase may be rotated away and we make this choice for $a_W$, while
keeping the rest complex.

A generic multi-doublet model, whether supersymmetric~\cite{sumrule} or
otherwise~\cite{Choudhury:2003ut}, is characterized by a sum rule for the
couplings of the neutral Higgs bosons to a pair of gauge bosons, namely $
\sum_i a^2_{VVH_i} = 1.  $ Although $a_{VVH_i}$ for a given Higgs boson can be
significantly smaller than the SM value (as, for example, may happen in the
MSSM), any violation of the above sum rule would indicate either the presence
of higher $SU(2)_L$ multiplets or more complicated symmetry breaking
structures (such as those within higher-dimensional
theories)~\cite{Choudhury:2003ut}.  On the other hand, such couplings may
appear either from higher order corrections to the vertex in a renormalisable
theory~\cite{kniehl:NPB} or from higher dimensional operators in an effective
theory~\cite{operator}.  The couplings $b_V$ and $\tbV$ can arise from the
terms such as $F_{\mu \nu} F^{\mu \nu} \Phi^\dagger \Phi$ or $F_{\mu \nu}
\tilde{F}^{\mu \nu} \Phi^\dagger \Phi$ where $\Phi$ is the usual Higgs
doublet, $F_{\mu \nu}$ is the field strength tensor and $\tilde{F}_{\mu \nu}$
its dual \cite{operator}. The effects of still higher dimensional terms in the
trilinear vertices of interest can be absorbed into $b_V$ or $\tbV$ by
ascribing them with non-trivial momentum dependences. Clearly if the cut-off
scale $\Lambda$ of this theory is much larger than the typical energy at which
the scattering experiment is to be performed, the said dependence would be
weak. Thus, the momentum dependence of the form factors have a rather minor
role to play at the first generation linear colliders, especially for
$\Lambda \sim 1$ TeV.

Finally, note that, unlike in the case of the $ZZH$ couplings, the various
terms in the $WWH$ effective vertex can be ascribed definite properties (see
Table \ref{table:CPT Properties}) under each of the discrete transformations
$C$, $P$ and $\hat T$ where $\hat T$ stands for the pseudo-time reversal
transformation, one which reverses particle momenta and spins but does not
interchange initial and final states. That the imaginary parts of $b_W$ and
$\beta_W$ may lead to $CP \hat T$-odd observable is, of course, to be expected.

\begin{table}
\begin{center}
\begin{tabular}{|c|c|c|c|c|c|}
\hline Trans. & $a_W$ & $\mathcal R(b_W)$ & $\mathcal R(\tbw)$ & $\mathcal
 I(b_W)$ & $\mathcal I(\tbw)$\\ \hline 
$C$ & + & + & + & + & + \\
 $P$ & + & + & $-$ & $+$ & $-$ \\ 
$\hat T$ & + & + & $-$ & $-$ & $+$ \\ 
\hline
\end{tabular}
\end{center}
\caption{\em Transformation properties of the terms in the Lagrangian 
corresponding to the various couplings.}
\label{table:CPT Properties} 
\end{table}
\subsection{The process and cross sections}
\begin{figure}[htb]
\vspace*{-8ex}
\hspace*{1cm}
\input{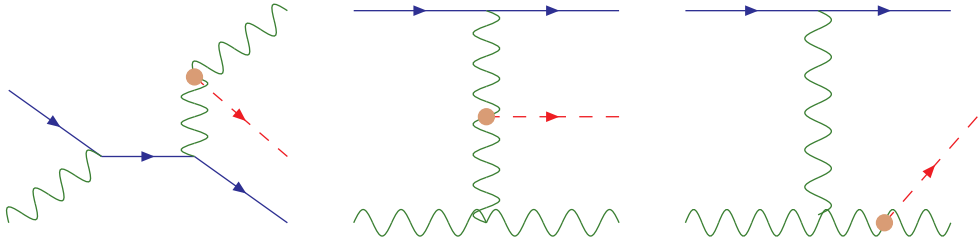}
\vskip 2.5cm
\caption{\em Feynman Diagrams for $e^- \gamma \to \nu_e W^- H$.}
  \label{Fig:feyn}
\end{figure}

To the lowest order, Higgs production at an $e\gamma$ collider--- the
process of Equation \ref{the_process}---receives contributions from the
three Feynman diagrams shown in Fig.\ref{Fig:feyn}.
In calculating the cross section, we retain  contributions  only
upto the lowest non-trivial order in the anomalous couplings, keeping
in view the higher dimensional nature of their origin.
Thus the cross-section
may be written as
\begin{equation}
\sigma = (1+ 2\; \Delta a_W)\sigma_0 +\mathcal R(b_W) \sigma_1 +
\mathcal R( \tbw) \sigma_2 + \mathcal I( b_W) \sigma_3 +\mathcal
I(\tbw) \sigma_4 .
\label{def_cross}
\end{equation}
Note that, as in Ref. \cite{bcgs}, we have assumed that we are 
dealing with a SM-like Higgs and hence 
\beq
    a_W \equiv 1 + \Delta a_W
\eeq
is close to its SM value. 

\begin{figure}[!h]
\vspace*{-21ex}
\epsfig{file=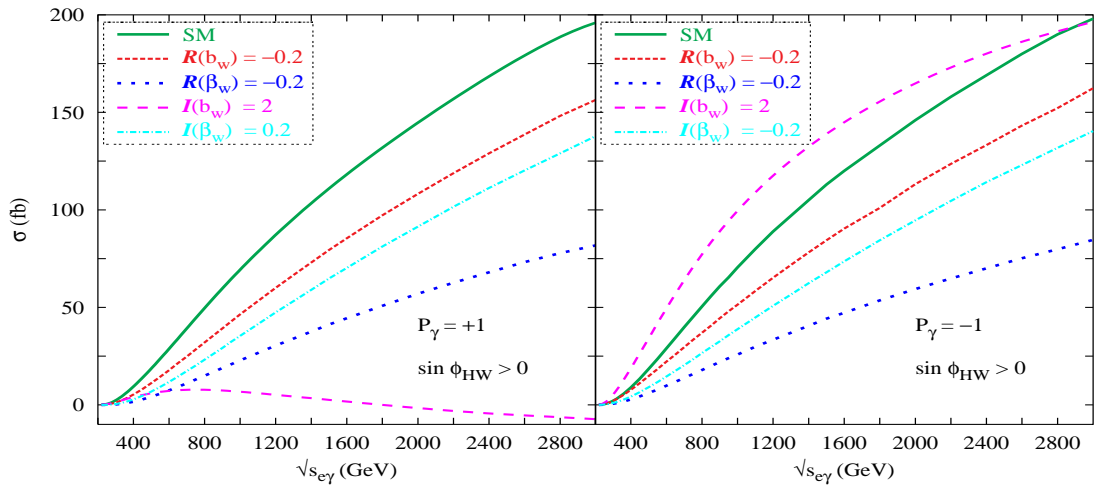,width=17cm,height=10cm}
\vspace*{-3ex}
\caption{\em Partial cross-sections $\sigma_i$ for 
$e^- \gamma \to \nu_e W^- H$, with the restriction that 
$\sin\phi_{HW} > 0$ as functions of the c.m. energy for a 
Higgs boson of mass 120 GeV and particular values for the 
anomalous couplings. Fully polarized but monoenergetic 
photons have been used, with the left(right) panels corresponding to 
$P_{\gamma} = +1 \, (-1)$ respectively. The electron is assumed to be 
unpolarized.}
\label{fig:rts_var} 
\end{figure}

Being odd under $\hat T$, some of the terms in
Equation (\ref{vertex}) would not contribute, at the linear order, to the
total rate, which is a $\hat T$ even observable. Thus, to see
their effect, we need to restrict ourselves to an appropriate
part of the phase space.  To this end, consider a frame wherein the
initial state $e^-$ points along the positive $z$-axis and defines,
along with the Higgs momentum, the $x-z$ plane.  An appropriate phase
space choice is described by a restriction of the $W$ to be produced
either in the hemisphere above or below this $x-z$ plane, or, in other
words, restricting $\sin \phi_{HW}$ ($\phi_{HW}$ being the azimuthal
separation between the Higgs and the $W$) to either a positive or a
negative value.  With this constraint applied, the different
contributions to the total rate, as a function of the center-of-mass
energy is displayed in Fig.\ref{fig:rts_var}.  [Note that a non-zero
$\Delta a_W$ would only rescale the SM rates.]
\vskip .5cm

Clearly, the SM cross-section grows with $\sqrt s_{e\gamma}$ for the range
shown, a consequence of the presence of the ``$t$''- and ``$u$''-channel
diagrams. It can be checked easily that, for each polarization state, the
contributions corresponding to $\rbw$, $\rbetw$ and $\ibetw$ asymptotically
grow faster than the SM cross section $\sigma_0$. This is but a reflection of
the higher-dimensional nature of the couplings. The contribution proportional
to $\ibw$, namely $\sigma_3$, on the other hand, grows slower than
$\sigma_0$. While this may seem surprising at first, this owes itself to the
fact that we are considering only the interference terms with the SM and that
these typically suffer from at least the same suppressions as the SM piece
(this is quite akin to the case of Ref.\cite{bcgs}).  In addition, it should
be remembered that the unrestricted ({\em i.e.}, summed over the full phase
space) $\sigma_2$ and $\sigma_3$ vanish identically.

The preceding discussion also indicates that a larger
$\sqrt{s_{e\gamma}}$ would increase the sensitivity to almost all the
couplings, bar $\ibw$.  However, rather than investigate the advantage
of varying the center of mass energy, we shall, henceforth, restrict
ourselves to a realistic first generation photon collider and
emphasize the importance of imposing various kinematic restrictions.
Such a choice would  naturally restrict us to momentum transfers far
below the deemed scale of new physics, viz. $\Lambda \sim 1 \tev$,
thereby allowing us to neglect any form-factor behavior for these
couplings.

A further point to be noted is the large dependence of the $\ibetw$ 
and $\ibw$ contributions on the photon polarization, whereas 
the others have only a minor dependence. This, of course, 
implies that polarized scattering may be used effectively to isolate 
the first two, a possibility that we shall return to later.
\section{A Realistic Collider Experiment}
\label{sec:real_expt}
\subsection{The photon collider}
Although we have considered a monoenergetic photon beam in deriving the
results in the last section, in reality, a high energy monochromatic
beam is not possible. Rather, a high energy photon beam is to be
obtained by back-scattering a laser beam from an electron/positron
beam. The reflected photon beam carries off only a fraction ($y$) of
the $e^\pm$ energy with 
\beq 
\barr{rcl} 
y_{\rm max} & = & \dis
\frac{z}{1 + z} \\[2ex] 
z & \equiv & \dis \frac{4 E_b E_L}{m_e^2} \; \cos^2 \frac{\theta_{b L}}{2} \ , 
\earr 
\eeq 
where $E_{b (L)}$ are the energies of the incident electron (or positron)
beam and the laser respectively and $\theta_{b L}$ is the incidence
angle.  In principle, one can increase the photon energy by increasing
the energy of the laser beam. However, a large $E_L$ (or,
equivalently, a large $z$) also enhances the probability of electron
positron pair creation through laser and scattered-photon
interactions, and consequently results in beam degradation.  An
optimal choice is $z = 2(1 + \sqrt{2})$, and this is the value that we
adopt in our analysis.

The cross-sections for a realistic electron-photon collider can then
be obtained by convoluting the fixed-energy cross-sections
$\hat{\sigma}({\hat s}, P_{\gamma}, P_{e^-})$ with the appropriate
photon spectrum:
\beq
\sigma(s) 
= \int {\rm d} y \; {\rm d} \hat s \; \:
    \frac{{\rm d} n}{{\rm d} y} (P_{b}, P_L) \; \:
\hat{\sigma}({\hat s}, P_{\gamma}, P_{e^-}) \; \delta(\hat s - y s) \ ,
        \label{csec_convolution}
\eeq
where the photon polarization is itself a function of $P_{b, L}$ and
the momentum fraction, viz. $P_{\gamma} = P_{\gamma}(y, P_b,
P_L)$. For simplicity, we shall consider only circularly polarized
lasers scattering off polarized electron (positron) beams. The
corresponding number-density $n(y)$ and average helicity for the
scattered photons are then given by~\cite{telnov}
\beq
\barr{rcl}
\dis \frac{dn}{dy} &=&  \dis 
        \frac{2 \pi \alpha^2}{m_{e}^2 z \sigma_C} C(y) 
   \\[2ex]
P_\gamma (y) &=& \dis 
        \frac{1}{C(y)} \bigg[ P_b \bigg\{ \frac{y}{1-y} + y(2 r -1)^2
\bigg \} - P_L (2 r -1) \bigg( 1 - y + \frac{1}{1-y} \bigg) \bigg] 
    \\[2ex]
C(y) &\equiv& \dis 
        \frac{y}{1-y} + (1 -y) - 4r(1-r) - 2P_b P_L rz (2r -1)(2 -y) \ ,
\earr
        \label{photon_spectrum}
\eeq
where $r \equiv y / z / (1 - y)$ and the total 
Compton cross-section $\sigma_C$ provides the normalization.
\subsection{The final state}
\label{subsec:kinem_cuts}

To be quantitative, we shall choose to work with a Higgs
boson of mass $120 \gev$ and a parent $e^+ e^-$ machine
operating at a center of mass energy of $500 \gev$, or, in other words
a maximum $\sqrt s_{e \gamma}$ of $\sim 455 \gev$. For such a Higgs 
mass, the dominant decay channel is the one into a $b \bar b$ pair, with 
the corresponding  branching fraction $\sim 0.9$. And as we want the 
$W$ momentum to be reconstructible, we restrict ourselves to the $W
\longrightarrow q \bar q$ mode with a branching 
fraction $\sim 0.68$. The final state thus comprises of four jets 
and missing momentum. Of the jets, two must be $b$-like and these 
must reconstruct to $m_H$ and the other two must reconstruct 
to $m_W$. 

To be detectable, each of the jets must have a minimum energy and 
they must not be too forward or backward.  Furthermore, any two
jets should be well separated so as to be recognizable as separate
ones. And finally, the events must be characterized by 
a minimum missing transverse momentum. 
To be quantitative, our acceptance cuts constitute 
\beq
\barr{rclcl}
p_T^{\rm miss} & \geq & 20 \gev & & \\
-3.0 \,\leq \,\eta_j & \leq & 3.0  & \qquad & \textup{for rapidity of
  each jet}  \\ 
p_T & \geq & 10 \gev & & 
\textup{for each jet}  \\ 
\Delta R_{j_1 j_2} & \geq & 0.7 & & 
\textup{for each pair of jets} 
\earr
\label{accept_cuts} 
\eeq
where $(\Delta R_{j_1 j_2})^2 \equiv (\Delta \phi)^2 + (\Delta
\eta)^2$ with $\Delta \phi$ and $\Delta \eta$ denoting the
separation between the two jets in azimuthal angle and rapidity
respectively.  

With only the acceptance cuts in place, and with the use of unpolarized 
beams (electron, laser as well the beam reflected off), the 
cross section is 
\begin{equation}
 \sigma = \left[ 4.15 \,(1 + 2\; \Delta a_W) - 16.09 \; \mathcal R (b_W) -
 1.96 \;\mathcal I (\tbw) \right] \, \fb \ .
\label{basic_cross_unpol}
\end{equation}
As expected, $\mathcal I (b_W)$ and $\mathcal R (\tbw)$ do not contribute
to the total rate, while $\mathcal I (\tbw)$ makes only a 
small contribution. 

For any such measurement, one may define a 
statistical measure of agreement with the SM expectations by 
defining a fluctuation through 
\begin{equation}
 (\delta \sigma )^2= \frac{\sigma_{_{SM}}}{{\mathcal L} +\epsilon^2
 \;\sigma^2_{_{SM}}}. 
\label{del_sig}
\end{equation}
Here $\sigma_{_{SM}}$ is the SM value of the cross-section, $ \mathcal
L$ is the integrated luminosity of the machine and $\epsilon$ is the
fractional systematic error. We shall, henceforth, consider $ \epsilon = 0.01$.
Using the total cross sections ---Equation \ref{basic_cross_unpol} ---alone,  
we can then constrain a particular linear combination of couplings, {\em viz.}
\begin{equation}
\mid 2 \;\Delta a_W - 3.88 \;\mathcal R (b_W) - 0.47 \; \mathcal I
(\tbw)\mid \;\leq \;0.072 \ .
\label{lim_basic_unpol}
\end{equation} 
at the $3 \sigma$ level.

\begin{figure}[!h]
\epsfig{file=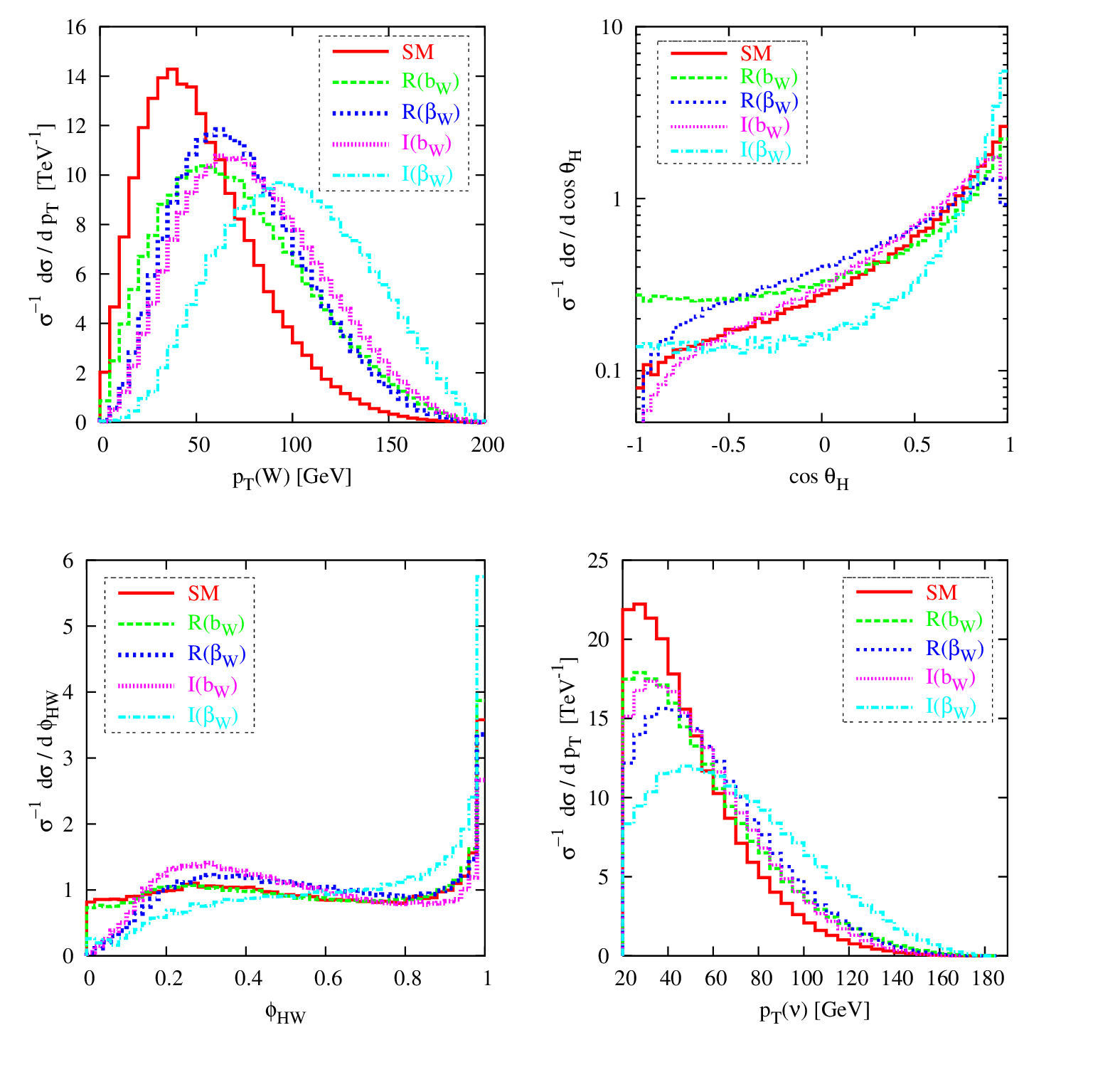,width=17.5cm,height=9cm}
\vspace*{-5ex}
\caption{\em Various kinematical distributions for unpolarized
scattering with the acceptance cuts augmented by the requirement of
$\sin\phi_{HW} > 0$.}
\label{fig:dist}
\end{figure}

As can be well appreciated, total cross sections are unlikely to be the
most sensitive of probes. For one, this observable is not at all sensitive
to either of $\mathcal I (b_W)$ and $\mathcal R (\tbw)$. Secondly, 
it is quite conceivable that contributions proportional to different 
anomalous pieces have distinct phase space distributions, thereby 
affording us the possibility of relative enhancement by choosing 
appropriate kinematical constraints. In Fig.\ref{fig:dist}, we display, 
for unpolarized scattering, some of the distributions 
wherein the differences are more prominent. 
%
\section{Results and Discussions}
\subsection{Unpolarized Beams}
\label{sec:unpol}

Although, as Fig.\ref{fig:rts_var} demonstrates, some of 
the anomalous contributions to the cross section have a strong dependence 
on the beam polarization, we refrain from using this at the outset, and 
confine ourselves to unpolarized scattering. Instead, we concentrate initially 
on devising appropriate phase-space restrictions so as to enhance a given 
contribution with respect to the SM one as also the others. Clearly, 
the former objective cannot be reached for the contribution proportional 
to $\Delta a_W$ as it is identical to the SM one in all the phase-space 
distributions. Hence, a measurement of $\Delta a_W$ has, of necessity, 
to be based on a counting measurement.

As has already been mentioned, the $\hat T$-odd couplings $\ibw$ 
and $\rbetw$ do not contribute to the total cross section, or even 
to partial ones as long as the phase space is $\hat T$-even. This, then,
constitutes a simple method of eliminating them from the analysis thereby 
allowing us to concentrate on the other three.

\subsubsection{ $\hat T$ even Couplings $\Delta a_W, \mathcal R (b_W)$ and 
$\mathcal I (\tbw)$ }
\label{subsec:unpol_T-even}

As a perusal of Fig.\ref{fig:dist} shows, the differential distributions for
the $\Delta a_W$ and $\mathcal R (b_W)$ contributions are not very dissimilar
and, hence, it is difficult to separate these two effects. The relative
contribution of $\mathcal I (\tbw)$, on the other hand, can be enhanced or
reduced upon the use of different cuts on kinematic observables. A partial
list of such cuts and the corresponding cross sections is displayed in Table
\ref{rts_500_unpol}.
\begin{table}[!h]
\begin{center}
\begin{tabular}{|l|l|c|r|r|}
\hline 
\multicolumn{2}{|c|}{\bf Cut} & {\textbf{ ${ \sigma_0}$ }}& {\textbf{$ \sigma_1$}}
& {\textbf{ $ \sigma_4$ }}\\ 
\hline 
$\mathcal C_0$ & { Acceptance cuts} & {
$4.15$ }& { $-16.10 $} & { $-1.96 $}\\
\hline
$\mathcal C_1$ &
 {  $p_{_T}(W) \geq 80$ GeV and } & {
 $0.25 $} & { $- 2.58 $} & { $- 0.73 $} \\
& { $\mid \sin{\phi_{HW}} \mid \geq 0.4$} & & &\\

\hline
$\mathcal C_2$ & { $p_{_T}(W) \geq 80$ GeV and } & {  $ 0.19 
$ }& { $ - 2.37 $} & { $ -0.74 $} \\
& { $p_{_T}^\textup{miss} \geq 60$ GeV} &  & & \\

\hline
$\mathcal C_3$ & {  $p_{_T}(W) \leq 80\;$ GeV and } & { $
1.11 $} & { $ - 2.55$ } & { $ 0.18 $} \\
& { $\mid \sin{\phi_{HW}} \mid \leq 0.4$} & & &\\
\hline
$\mathcal C_4$ & { $p_{_T}(W) \leq 80$ GeV and }  & { $ 1.89 
$ } & { $ - 5.56  $} & { $ 0.044 $}  \\
&  { $ \mid \cos{\theta_{_H}} \mid \leq 0.8$} & & & \\
\hline
$\mathcal C_5$ & { $p_{_T}(W) \geq 80\;$ GeV and } & { $
0.50 $} & { $ - 3.49$ } & { $ -0.62 $} \\
& { $\mid \sin{\phi_{HW}} \mid \leq 0.4$} & & &\\
\hline
\end{tabular}
\end{center}
\caption{\em Various cuts and the corresponding rates, in femtobarns, for
unpolarized scattering with $\sqrt {s_{ee}} = 500$ GeV.}
\label{rts_500_unpol}
\end{table}

The set of cuts $\mathcal C_3$, and even more convincingly, $\mathcal
C_4$, eliminates the bulk of the $\mathcal I(\tbw)$ contribution. Assuming,
for the moment, that the anomalous couplings are of the same order,
the imposition of such cuts would allow us to neglect the presence of
even a non-zero $\mathcal I(\tbw)$ and instead impose a constraint on
particular combinations of $ \Delta a_W$ and $\mathcal R (b_W)$.  For
example, with the use of cut $\mathcal C_3$, the rate depends on the
combination
\begin{equation}
\eta_3 =  2\; \Delta a_W -2.30\;  \mathcal R (b_W) \,\quad {\rm , viz }\quad
\sigma(\mathcal C_3) \approx \sigma_0\, (1 + \eta_3) = 1.11\, (1 + \eta_3)
\label{eta_3}
\end{equation}
and, thus, for an integrated luminosity of $500 \; {\rm fb}^{-1}$, the lack of
a deviation from the SM expectation values would give us 
a $3\,\sigma $ limit on $\eta_3$, namely 
\begin{equation}
  \mid \eta_3 \mid \;\leq \; 0.13 \ .
\label{eta_3_limit}
\end{equation}
Similarly, the use of $\mathcal C_4$ results in 
\begin{equation}
\sigma(\mathcal C_4) = \sigma_0\, (1 + \eta_4) = 1.89\, (1 + \eta_4) \, , {\rm
with} \qquad \eta_4 = 2\; \Delta a_W - 2.94\; \mathcal R (b_W).
\end{equation}
This results in $3\,\sigma $ bound on $\eta_4$ of 
\begin{equation}
 \mid
 \eta_4 \mid \;\leq \; 0.10.
\label{eta_4_limits}
\end{equation}

Contrary to $ \mathcal C_{3,4}$, the cuts $ \mathcal C_{1,2}$ serve to
enhance the effect of $\ibetw$, though not to the extent that the
effects of the other two may be entirely neglected.  Consequently, we
can constrain only certain linear combinations of the three, viz.
\beq
\barr{rclcl}
\mid \;2\;\Delta a_W - 10.36 \;\mathcal R (b_W) - 2.93 \; \ibetw \mid &
 \leq & \;0.27  & \qquad & ({\rm using \ cut}\;\; \mathcal C_1)  \\[1ex]
\mid \;2\;\Delta a_W - 12.25 \;\mathcal R (b_W) - 3.84 \; \ibetw \mid \;
 & \leq & \;0.30 & &  ({\rm using \ cut}\; \; \mathcal C_2).
\label{lim_c1_c3}
\earr \eeq If we make the simplifying assumption that only one anomalous
coupling may be non-zero, the corresponding limits are easy to obtain. The
strongest such limits are listed in Table \ref{ind_limits_re_unpol}.
\begin{table}[!h]
\begin{center}
\begin{tabular}{|c|c|c|}
\hline
Coupling & $ 3 \sigma$ bound & Observable Used \\
\hline
$\mid \Delta a_W \mid$ &  0.050 & $\sigma$ with $\mathcal C_4$ \\
$\mid \mathcal R(b_W) \mid$  & 0.035 & $\sigma$ with $\mathcal C_4$ \\
$\mid \ibetw \mid$  & 0.078 & $\sigma$ with $\mathcal C_2$ \\
\hline
\end{tabular}
\end{center}
\caption{\em Achievable upper limits ($3 \sigma$) on $ \Delta a_W$, $\mathcal
R (b_W)$ and $\ibetw$, under the assumption that only one of the couplings is
non-zero. An integrated luminosity of $500\; {\rm fb}^{-1}$ using unpolarized
beams has been assumed.}
\label{ind_limits_re_unpol}
\end{table}

\begin{figure}[!h]
\vspace*{-35ex}
\hspace*{-1.5cm}
\epsfig{file=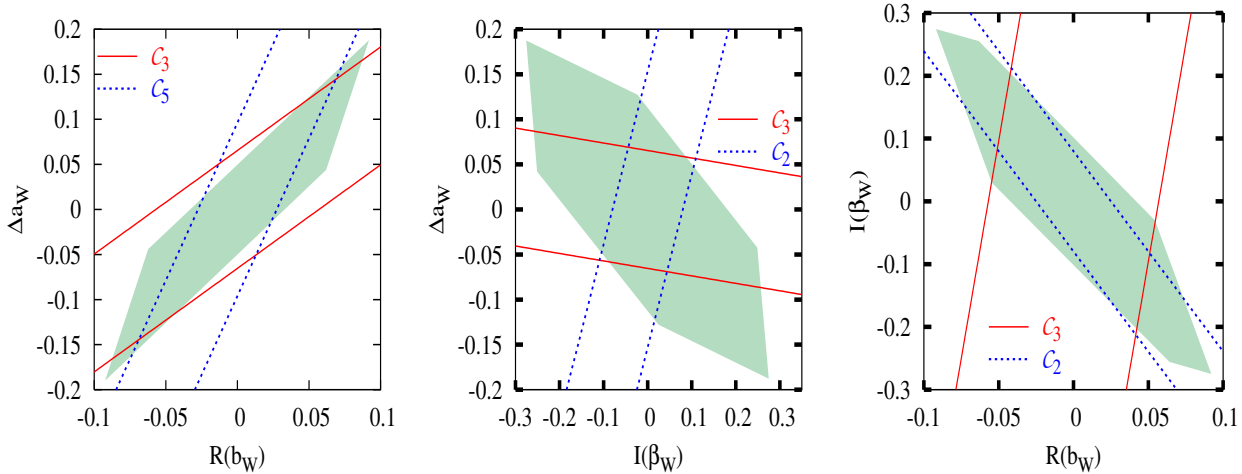,width=\textwidth, height=0.9\textwidth}
\vspace*{-5ex}
\caption{\em The pairs of oblique lines  denote the region allowed by the
  corresponding cut, at the $3\, \sigma$ level, when the third anomalous
  coupling is identically zero. Intersection of strips, thus, gives the area
  allowed by both the observables. The shaded regions constitute the
  projections of the parameter space that leads to observables
  indistinguishable from the SM expectations for each of the cuts of Table
  \protect\ref{rts_500_unpol} when all three couplings are allowed to be
  non-zero. In each case, an integrated luminosity of 500 fb$^{-1}$ has been
  used. }
\label{fig:constr_unpol}
\end{figure}

While individual limits might be strong and interesting in their own right, it
is of importance to investigate how well the couplings may be resolved. A
simple way would be to consider two of the couplings at a time (assuming the
third to be vanishing) and impose different constraints on such a plane. As we
are working in the linear approximation, all such constraints, for a given
observable, would naturally result in an infinite linear strip as the allowed
region. The intersection of such strips for mutually exclusive observables
would, then, constitute the region of interest.  In
Fig.\ref{fig:constr_unpol}, we display this for each of the three combinations
of anomalous couplings.
A more general analysis keeping all the three couplings non-zero is possible,
and quite straightforward. Demanding that, for a combination of such couplings
to be allowed, the observables corresponding to each of the cuts of Table
\ref{rts_500_unpol} must be indistinguishable from the SM expectations, we
generate a three-dimensional volume of the allowed parameter space. In
Fig.\ref{fig:constr_unpol}, we project such a $3 \sigma$ volume onto the three
different planes. It is reassuring to note that a non-zero value of $\ibetw$
has relatively little role to play in the constraints in the $\Delta
a_W$--$\rbw$ plane, thereby vindicating our intermediate approximation of
neglecting this contribution.  This, of course, is just a consequence of the
smallness of the $\ibetw$ contribution once ${\cal C}_3$ is imposed.
Analogous features are displayed by the projections on the $\Delta
a_W$--$\ibetw$ and $\rbw$--$\ibetw$ planes as well. The simultaneous limits
obtained from these shaded regions of Fig. \ref{fig:constr_unpol} are given in
Table \ref{sim_limits_unpol}.
\begin{table}[!h]
\vskip 1cm
\begin{center}
\begin{tabular}{|c|c|}
\hline Coupling & $ 3\; \sigma$ bound  \\ 
\hline 
$\mid \Delta a_W \mid$ & 0.19 \\ 
$\mid \mathcal R(b_W) \mid$ & 0.092   \\
 $\mid \mathcal I(\tbw) \mid$ & 0.27   \\ 
\hline
\end{tabular}
\end{center}
\caption{\em Simultaneous limits on anomalous couplings at $3 \sigma$ level,
  based on shaded regions of Fig. \ref{fig:constr_unpol}, using unpolarized
  beams for an integrated luminosity of $500 \, {\rm fb}^{-1}$.}
\label{sim_limits_unpol}
\end{table}
\subsubsection{ $\hat T$ odd Couplings $ \mathcal I (b_W)$ and 
$\mathcal R (\tbw)$ }
\label{subsec:unpol_T-odd}
As already described above, the imaginary parts of $b_W$ and $\beta_W$
do not contribute to the total rate on account of being odd under
naive time-reversal.  However, on restricting the $W$-boson to lie above
the plane of production of the Higgs (defined in conjunction with the
beam axis), \textit{i.e.}  on requiring $\sin{ \phi_{HW}} \geq 0$, we
have, for the corresponding partial cross-section,
\begin{equation}
\barr{rcl}
 \sigma (\sin{ \phi_{HW}} \geq 0) & = & 2.07 \;(1 + 2\; \Delta a_W) -
8.04 \; \mathcal R (b_W) - 0.982 \; \ibetw
\\
& + & 
1.50 \; \mathcal I (b_W) - 3.11 \; \mathcal R (\tbw).
\earr
\label{cs_unpol_sin_posit}
\end{equation}
For events in the other hemisphere ($\sin{ \phi_{HW}} \leq 0$), the
contributions corresponding to $\Delta a_W$, $\rbw$ and $\ibetw$
(i.e. $\sigma_{0,1,4}$) remain the same, while the contributions corresponding
to couplings $\rbetw$ and $\ibw$  ($\sigma_{2, 3}$) reverse sign.  This, then,
prompts the use of a $\hat T$-odd asymmetry of the form
\begin{equation}
\barr{rcl}
 \mathcal A & \equiv & \dis {\frac{\sigma_{\sin{ \phi_{HW}} \geq 0} -\sigma_{\sin{
 \phi_{HW}} \leq 0}}{\sigma_{\sin{ \phi_{HW}} \geq 0} +
 \sigma_{\sin{ \phi_{HW}} \leq 0}}}  \\ [3ex]
&= & \dis {\frac{ \left[3.0 \; \mathcal I(b_W) - 6.22
 \;\mathcal R (\tbw) \right]}{ 4.15\; \left[(1 + 2\;
 \Delta a_W) - 7.76 \; \mathcal R(b_W) - 0.946 \;\mathcal I (\tbw) \right]}}  \ ,
\earr
\label{asymm_unpol}
\end{equation}
with the corresponding fluctuation in the measurement being given by 
\begin{equation}
(\delta \mathcal A)^2 = \frac{1- \mathcal A_{SM}^2}{\sigma_{SM} \;
\mathcal L}+ \frac{\epsilon^2}{ 2} \; (1- \mathcal A_{SM}^2)^2 \ .
\label{delta_A}
\end{equation}
It should be noted that the asymmetry vanishes identically within the SM. 
A glance at the various distributions of Fig.\ref{fig:dist} shows that
the employment of further kinematic cuts do not alter the relative
contributions of $ \mathcal I(b_W)$ and $\mathcal R(\tbw)$ in any
significant way.  Thus, with unpolarized beams, the best bounds on
these two couplings are obtained from Equation \ref{asymm_unpol} and, for
an integrated luminosity of $500 \; fb^{-1}$, reads
\begin{equation}
\mid 1.50\; \mathcal I(b_W) - 3.11 \;\mathcal R (\tbw) \mid
\; \leq 0.14
\label{imbw_imbwt with sm denom}
\end{equation}
at the $3 \sigma$ level.  Note that, in deriving the above, we have
neglected the anomalous contributions in the denominator of
Equation(\ref{asymm_unpol}), which is in consonance with our approximation of 
retaining terms which are at best linear in the anomalous couplings.
Once again, if we assume that only one of these is non-zero, the corresponding 
$3 \sigma$ bounds are 
\beq
\mid \mathcal I(b_W) \mid \leq  0.092  \qquad {\rm and} \qquad
\mid \mathcal R(\tbw) \mid \leq  0.045 \ .
\label{ind_lim_T-odd_unpol}
\eeq
\subsection{Polarized Beams}
\label{sec:pol}
Having exhausted the possible ways that unpolarized $e^- \gamma$
scattering could be used to probe the $WWH$ vertex, we now examine the
role, if any, of beam polarization. Since the dependence on $e^-$
polarization is trivial (only $e^-_L$ contributes to the process under
consideration), we choose to neglect this, while noting that having a
left-polarized electron will only serve to rescale both the signal and
the background in an identical fashion, thereby improving the
statistical significance. Concentrating on the non-trivial dependence
of the cross section on the photon polarization, we note that the
latter is a function, vide Equation (\ref{photon_spectrum}), of the beam
and laser polarizations.  While the laser can be polarized fully, the
beam polarization is unlikely to be much higher than 80\%. We, thus,
choose four different combinations, namely
\begin{eqnarray}
a : \quad (P_L, P_b) = (+1, +0.8)\nonumber \\
b : \quad (P_L, P_b) = (-1, -0.8)\nonumber \\
c : \quad (P_L, P_b) = (+1, -0.8)\nonumber \\
d : \quad (P_L, P_b) = (-1, +0.8)
\label{pol_comb}
\end{eqnarray}
and associate a given mode with one-fourth of the total integrated
luminosity, viz. $125 \fb^{-1}$ each.  The corresponding rates, on
imposition of the acceptance cuts alone are given in
Table~\ref{pol_T-even_cs}.
%
%
\begin{table}[!h]
\begin{center}
\begin{tabular}{|l|l|c|c|r|}
\hline 
\multicolumn{2}{|c|}{\textbf{$ (P_L, P_b)$} } &  {\textbf{$ \sigma_0$}} &
{\textbf{$ \sigma_1$}} & {\textbf{ $\sigma_4$ }}\\ 
\hline
$\sigma_{P a}$ & {\small $(+1, +0.8)$}   &{\small $3.26 $}& {\small $-11.28$} & {\small $2.88$}\\

\hline
$\sigma_{P b}$ & {\small $(-1, -0.8)$ }  & { \small $ 3.47 $}& {\small   $- 15.25 $}& {\small $- 6.27 $}\\ 
\hline
$\sigma_{P  c}$ &  {\small $(+1, -0.8)$}  &  {\small $5.13 $} &   {\small $- 21.16 $} & {\small $-9.97 $ }\\ 
\hline
 $\sigma_{P d}$ &  {\small $(-1, +0.8)$}  & { \small $4.96  $} & {\small $ - 16.95 $} & {\small $ 5.44  $} \\
\hline 
\end{tabular}
\end{center}
\caption{\em Cross-sections for various polarization combinations  with only
  acceptance cuts imposed.}
\label{pol_T-even_cs}
\end{table}
As could have been easily surmised from Fig.\ref{fig:rts_var}, of the
five cross sections $\sigma_{0 \dots 4}$, polarization dependence is
maximal for that proportional to $\ibetw$ followed by that for the
$\ibw$ term. Of course, the latter contribution vanishes identically
if integrated over a symmetric phase-space. As discussed before,
imposition of various cuts may enhance or reduce the relative
contributions of different anomalous couplings. Of the four
polarization combinations of Equation (\ref{pol_comb}), the first one
($a$) was not found to be significantly useful. The cross-sections
obtained after imposing further cuts on the other three 
combinations of polarization are given in Table \ref{cs_pol_cuts}.

\begin{figure}[!h]
\vspace*{-40ex}
\hspace*{-1.5cm}
\epsfig{file=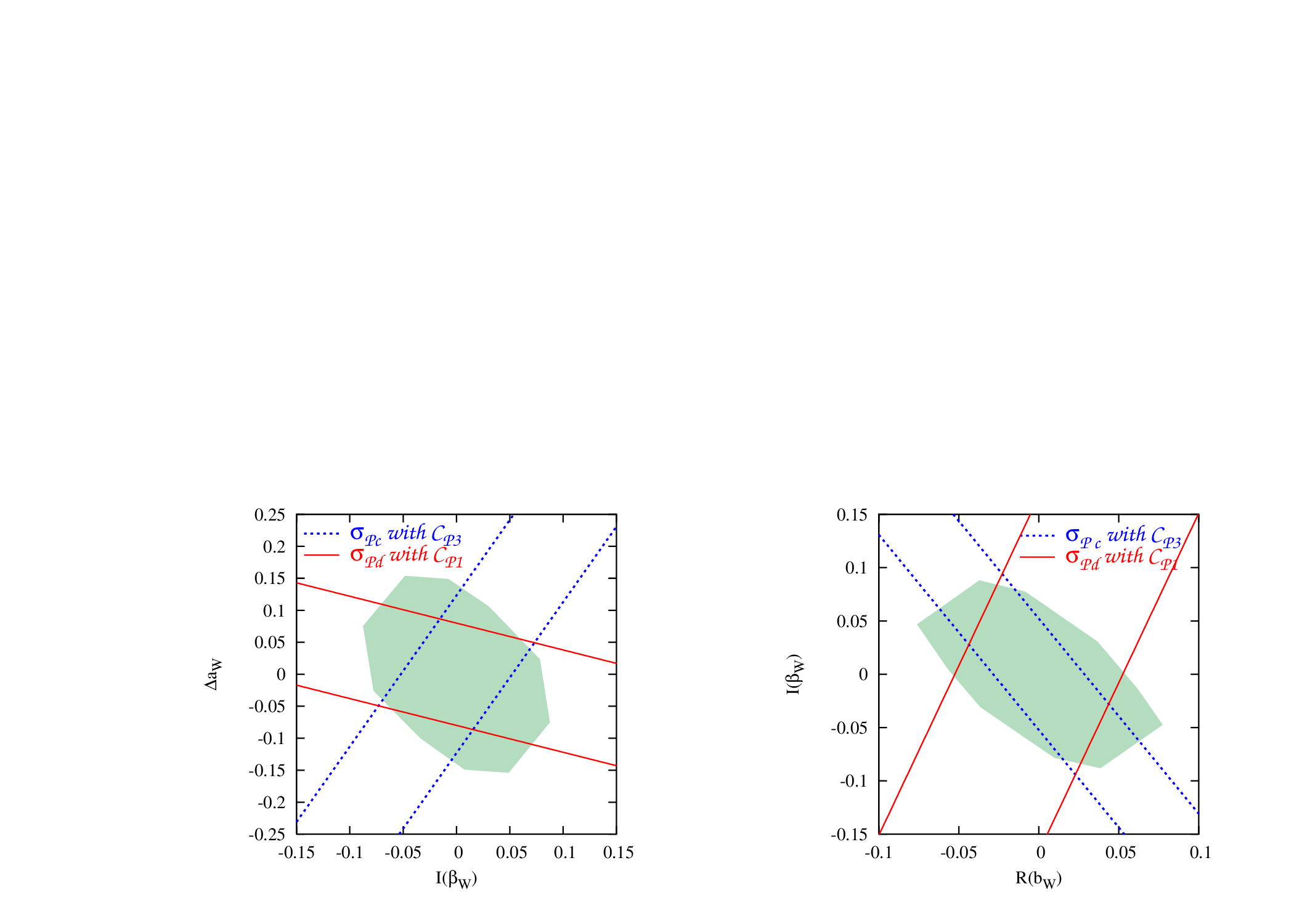,width=\textwidth, height=0.9\textwidth}
\vspace*{-5ex}
\caption{\em The allowed regions in the $\mathcal I(\beta_W)-\Delta
a_W$ and $\rbw-\ibetw$ planes. The shaded regions constitute the
projections of the parameter space that leads to observables
indistinguishable, at the $3\, \sigma$ level, from the SM expectations
for each of the cuts given in Table \protect\ref{cs_pol_cuts} and an
integrated luminosity of $125 \, {\rm fb}^{-1}$. The pairs of oblique
lines denote the region allowed by the corresponding cut when the
third anomalous coupling is identically zero.}
\label{fig:constr_T-even_pol}
\end{figure}

\begin{table}
\begin{center}
\begin{tabular}{|c|l|}
\hline
Cut name & \multicolumn{1}{c|}{Cut Description} \\
\hline
$\mathcal C_{P1}$ & $p_{_T}(W) \leq 75 $ GeV  \\
$\mathcal C_{P2}$ & $p_{_T}(W) \leq 75 $ GeV and $\cos{\theta_H} \leq 0 $  \\
$\mathcal C_{P3}$ & $p_{_T}(W) \geq 75 $ GeV  \\
$\mathcal C_{P4}$ & $p_{_T}(W) \geq 75 $ GeV  
          and $\cos{\theta_W} \geq 0 $ \\
\hline
\end{tabular}
\vskip 10pt
\begin{tabular}{|c||c|c|c||c|c|c||c|c|c||}
\hline 
& \multicolumn{3}{c||}{$b: (P_L,\; P_b) = (-1,\;-.8)$}
& \multicolumn{3}{c||}{$c: (P_L,\; P_b)= (+1, \; -.8)$}
& \multicolumn{3}{c||}{$d: (P_L,\; P_b)= (-1, \; +.8)$}
\\
\cline{2-10}
\bf{ cut} & 
{\textbf{$ \sigma_0$}} &{\textbf{ $ \sigma_1$}} & {\textbf{$ \sigma_4$}} 
&
{\textbf{$ \sigma_0$}} &{\textbf{ $ \sigma_1$}} & {\textbf{$ \sigma_4$}} 
&
{\textbf{$ \sigma_0$}} &{\textbf{ $ \sigma_1$}} & {\textbf{$ \sigma_4$}} \\ 
\hline 
$\mathcal C_{P1}$ 
&  2.75 & $- 8.8$ & $ -3.16$ 
& 3.91 & $- 11.23$ & $ -4.31 $
& $ 3.85$ &$ - 10.27$ & $ 3.23$ \\
\hline 
 $\mathcal C_{P2}$ & 
  0.41 & $-2.21$    & $- 0.10$
& & &  & & & \\ 
\hline 
  $\mathcal C_{P3}$
& $0.72$ & $ - 6.47$  &  $ - 3.10$
& $1.20$ & $ - 10.40$  &  $- 5.68$
&$ 1.09$ &  $- 6.67$  & $ 2.20$\\
\hline
 $\mathcal C_{P4}$ & 
 $0.08$ & $-1.83 $   & $- 0.87$
& & & & & &  \\ 
\hline
\end{tabular}
\end{center}
\caption{\em The description of cuts over and above acceptance cuts for the study with polarized beams
and the corresponding cross-sections (in femtobarns).}
\label{cs_pol_cuts}
\end{table}
As a non-zero $\Delta a_W$ results in just a rescaling of the SM cross
section, the use of polarized beam naturally does not lead to any
significant improvement in its determination.  However, the relative
contributions of $\rbw$ and $\ibetw$ can be enhanced by imposing various
cuts. The strongest limits are derived using the cut $\mathcal C_{P3}$ on $\sigma_{Pc}$ ($P_T(W) \geq 75{\rm GeV\, for} \, (P_L, P_b) = (+1, -.8)$) which gives
\begin{equation}
\mid 2 \;\Delta a_W - 8.67 \;\mathcal R (b_W) -4.73\; \mathcal I
(\tbw)\mid \leq \; 0.25 .
\label{plaser-1_pbeam-.8_ptwgt70_cth_w+}
\end{equation}
Once again, on assumption of only one of these couplings being non-zero leads to
\begin{eqnarray}
\mid \mathcal R (b_W) \mid & \leq \; 0.029 \nonumber\\
\mid \mathcal I (\tbw) \mid & \leq \; 0.053 . \\
\label{ind_limit_rbw_rbwt_pol2}
\end{eqnarray}
These limits should be compared to those of Table
\ref{ind_limits_re_unpol}. It is to be noted that these, stronger, limits have
been derived using an integrated luminosity of only $125\, {\rm fb}^{-1}$.
Since the couplings cannot be isolated, we should be looking at the
simultaneous limits which are to be obtained from the graphs (following the
way it was done in section \ref{subsec:unpol_T-even} for the unpolarized
case). It should be noted from Table \ref{pol_T-even_cs} that the
contribution of $\ibetw$ has very strong dependence on the value of $(P_L, \; P_b)$. This fact is exploited to obtain constraints in the
$\Delta \, a_W - \ibetw$ and $\rbw - \ibetw$ planes. These simultaneous
constraints are shown in Fig. \ref{fig:constr_T-even_pol} and the limits
obtained are listed in Table \ref{sim_limits_pol}. It should be noted that the
combination  ($a$) of polarization {\it i.e.} $(P_L, P_b) =(+1, +0.8) $ is not
a very sensitive probe and hence it has been disregarded in obtaining these
simultaneous limits.
%

As before, to get contribution from the $\hat T$-odd couplings $ \mathcal
I (b_W)$ and $ \mathcal R (\tbw)$, we have to restrict ourselves
to only half of the phase space, namely, $\sin{ \phi_{HW}} \geq
0$. These partial cross-sections for various polarizations are given in Table 
\ref{cs_pol_sin_posit}.
\begin{table}[!b]
\begin{center}
\begin{tabular}{|l|l|c|r|r|c|c|}
\hline 
\multicolumn{2}{|c|}{ {\textbf{$ (P_L, P_b)$} }} & {\textbf{$ \sigma_0$}} &
{\textbf{$ \sigma_1$}} & {\textbf{$ \sigma_4$ }}& {\textbf{$ \sigma_3$}} &
{\textbf{$ \sigma_2$}}\\ 
\hline 
$ \sigma_{P a} $&{\small $(+1, +0.8)$} & {\small
$1.63$}& {\small $-5.64 $} & {\small $1.44$} & {\small $0.74$} &
{\small $ -2.03$}\\
\hline
$\sigma_{P b}$& {\small $(-1, -0.8)$ } & {\small $1.74 $}& {\small $-7.63 $} &
 {\small $-3.12 $} & {\small $1.75$} & {\small $-2.94$}\\
\hline
 $\sigma_{P c}$ & {\small $(+1, -0.8)$}   & {\small $2.56 $}& {\small $-10.82 $} & {\small $-4.98 $} & {\small $2.59$} & {\small $-4.43$}\\

\hline
$\sigma_{P d}$ & {\small $(-1, +0.8)$} & {\small $2.48 $}& {\small $-8.48 $} &
{\small $2.71 $} & {\small $0.99$} & {\small $-3.22$}\\ 
\hline
\end{tabular}
\end{center}
\caption{\em Cross-sections for polarized beams with acceptance cuts and with $\sin{ \phi_{HW}} \geq 0$.}
\label{cs_pol_sin_posit}
\end{table}

\begin{figure}[!h]
\vspace*{-15ex} \hspace*{6ex}
\epsfig{file=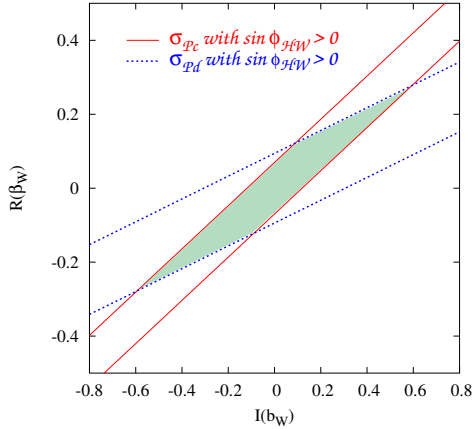,width=6.5cm}
\vspace*{-3ex}
\caption{\label{fig:constr_T-odd_pol} {\em The region in the $\mathcal
  I(b_W)-\mathcal R(\tbw)$ plane consistent with $3 \sigma$ variations in the
  assymetries with polarized photon beams for an integrated luminosity of 125
  fb$^{-1}$ per polarization combination. The pair of oblique lines correspond
  to ${\sigma_{P c}}$ and $\sigma_{P d}$ with $\sin{\phi_{HW}} \geq 0$ while
  the shaded region is obtained by demanding that the anomalous events can't
  be distinguished from SM values at $3 \sigma$ corresponding to
  all combinations of $(P_L, P_b)$ given in Table  \ref{cs_pol_sin_posit}.}}
\end{figure}

We construct the $\hat T$-odd asymmetry $\mathcal A$ as before and find that
the best limits are obtained for the case ($d$), namely $(P_L, Pb) = (-1,
+0.8) $. For this case,
\begin{equation}
 \mathcal A = \mathcal A_d = \frac{0.99\; \mathcal I(b_W) - 3.22 \;\mathcal R (\tbw)}{\;2.48 \left[(1 + 2\; \Delta a_W) - 3.42 \; \mathcal R(b_W) + 1.09
 \;\mathcal I (\tbw) \right]}\;. 
 \label{plaser-1_pbeam+.8_asymm}
\end{equation}
Using the cross-section for this case with only acceptance  cuts, namely
$\sigma_{Pd}$ from Table \ref{cs_pol_sin_posit}, we obtain
\begin{equation}
\mid 2 \;\Delta a_W - 3.42 \;\mathcal R (b_W) + 1.09 \; \mathcal I
(\tbw)\mid\, \leq 0.12 .
\label{plaser-1_pbeam+.8_basic}
\end{equation} 
Using Equation \ref{plaser-1_pbeam+.8_basic} and Equation
\ref{plaser-1_pbeam+.8_asymm}, the $3 \sigma$ bound on the asymmetry gives
us
\begin{equation}
\mid 0.99\; \mathcal I(b_W) - 3.22 \;\mathcal R (\tbw) \mid
\; \leq 0.14.
\label{lim_imbw_bwt_plaser-1_pbeam+.8}
\end{equation}
Keeping only one of these to be non-zero, we obtain the following individual
limits on them at  $3 \sigma$:
\begin{equation}
\mid \mathcal I(b_W) \; \mid \; \leq  0.15   \qquad {\rm and} \qquad
\mid \mathcal R(\tbw)\; \mid \; \leq  0.047 
\label{ind_limit_im_imt}
\end{equation}
for an integrated luminosity of $125 \; {\rm fb^{-1}}$.  Comparing with Equation
\ref{ind_lim_T-odd_unpol}, we observe that the individual limits are not
improved by use of polarized photon beam. This is because of reduction in
luminosity. However, using any pair of rates in Table \ref{cs_pol_sin_posit},
it is possible to obtain the allowed region in $\ibw - \rbetw $ plane which
can be used to put the simultaneous limits on these couplings. This was not
possible with the unpolarized photons. The constraints in $\mathcal I(b_W)
-\rbetw $ plane using $\sigma_{P c}$ and $\sigma_{P d}$ of Table
\ref{cs_pol_sin_posit} are given in Fig. \ref{fig:constr_T-odd_pol}. Since
the intersecting region given by two oblique lines and the shaded regions are
same, it is clear that the cases $a$ and $d$ ({\it i.e.} $ (P_L, P_b)
= (+1,\; +0.8)\;\; {\rm  and}\;\; (P_L, P_b) = (-1, \; -0.8) $) do not play any role in
isolating these two couplings.
We summarise in Table \ref{ind_limits_pol}, the individual limits on various
anomalous couplings and in Table \ref{sim_limits_pol} the simultaneous limits.
\begin{table}[!h]
\begin{center}
\begin{tabular}{|c|c|c|c|}
\hline 
Coupling & $ 3\; \sigma$ bound & Observable Used  \\ 
\hline
%
%
$\mid \mathcal R(b_W) \mid$ & 0.029 & $ \sigma_{Pc}$  with $\mathcal C_{P3}$\\ 
$\mid \mathcal I(\tbw) \mid$ & 0.053 & $ \sigma_{Pc}$ with $\mathcal
C_{P3}$\\
$\mid \mathcal I(b_W) \mid$ & 0.150 &  $\mathcal A_d$ \\
$\mid \mathcal R(\tbw) \mid$ & 0.047 & $\mathcal A_d$ \\ \hline
\end{tabular}
\end{center}
\caption{ \em Individual Limits on anomalous couplings at $3 \sigma$ level 
using polarized beams at an integrated luminosity of $125 \, {\rm fb}^{-1}$.}
\label{ind_limits_pol}
\end{table}
\begin{table}[!h]
\begin{center}
\begin{tabular}{|c|c|c|}
\hline Coupling & $ 3\; \sigma$ bound & Graph Used \\ 
\hline 
$\mid \Delta a_W \mid$ & 0.150 & Fig. \ref{fig:constr_T-even_pol}\\ 
$\mid \mathcal R(b_W) \mid$ & 0.078 & Fig. \ref{fig:constr_T-even_pol}  \\
 $\mid \mathcal I(\tbw) \mid$ & 0.088 &  Fig. \ref{fig:constr_T-even_pol}  \\ 
$\mid \mathcal I(b_W) \mid$ &  0.58 & Fig.
\ref{fig:constr_T-odd_pol}  \\
 $\mid \mathcal R(\tbw) \mid$ & 0.27  & Fig.
\ref{fig:constr_T-odd_pol}  \\ 
\hline
\end{tabular}
\end{center}
\caption{\em Simultaneous Limits on anomalous couplings at $3 \sigma$ level 
using polarized beams at an integrated luminosity of $125 \, {\rm fb}^{-1}$.}
\label{sim_limits_pol}
\end{table}
It should be noted that the simultaneous limits obtained are much better
compared to those obtained with unpolarized beams. This is particularly
apparent for the case of $\ibetw$. Also we are
able to obtain the simultaneous limits for $ \ibw$ and $\rbetw$. 
\section{ Use of the $e^+ \, \gamma$  initial state}
\label{sec:PT-conjugate}
Until now, in constructing observables, we have exploited only the
transformation of the operators of Equation \ref{vertex} under
either $P$ or $\hat T$. We now consider the action of the composite
discrete symmetry $P \hat T$. This is best achieved by comparing the
results obtained until now using the $e^- \, \gamma$ colliders with
those expected from the conjugate process, namely
 $$e^+ \; \gamma \lra \bar \nu \; W^+ \; H$$
\subsection{Unpolarized Beams}
\label{subsec:e+_unpol}
The total cross-section for the conjugate process is related to that obtained
earlier in a simple fashion:
\begin{eqnarray}
\sigma_{e^+ \gamma } &= &\left[\;\sigma_0\; (1 + 2\; \Delta a_W) + \sigma_1\;
 \mathcal R(b_W) + \sigma_2\; \mathcal R(\tbw) \right]_{e^- \gamma} \nonumber
 \\ 
& & - \left[  \sigma_3\; \mathcal I( b_W)+  \sigma_3\; \mathcal I( b_W)
 \right]_{e^- \gamma}. 
\label{rel_e+e-_cross_sec_unpol}
\end{eqnarray}
This leads us to  construct the asymmetry
\begin{eqnarray}
 \mathcal A_{C1_{\rm unpol }} &\equiv& \frac{\sigma_{e^+ \gamma }
-\sigma_{e^- \gamma }}{\sigma_{e^+ \gamma} +\sigma_{e^- \gamma}}
\nonumber \\
&=& \frac{2\; \left[\sigma_3 \;\mathcal I (b_W) +\sigma_4\; \mathcal I(\tbw)
 \right]_{e^+ \gamma }}{2\; \left[\sigma_0\; (1 + 2\; \Delta a_W) + \sigma_1\;
 \mathcal R(b_W)+ \sigma_2\; \mathcal R(\tbw) \right]_{e^+ \gamma} } \nonumber
 \\ 
&\simeq& \frac{1}{\sigma_0} \left[\sigma_3 \;\mathcal I (b_W) +\sigma_4\;
 \mathcal I(\tbw) \right]_{e^+ \gamma },
\label{asymm_Ac1_unpol}
\end{eqnarray}
where the approximate equality follows from our premise of retaining only
 terms linear in the anomalous couplings.  If no cut is imposed on
 $\sin{\phi_{HW}}$, the numerator contains only the $\mathcal I (\tbw)$ term
 allowing us to obtain a direct bound on his coupling alone, something that we
 were hitherto unable to.  With only the acceptance cuts imposed, we obtain
 for the $3 \sigma$ bound
\begin{equation}
 \mid \mathcal I(\tbw) \mid \;\leq  \; 0.14,
\label{asymm_lim_rbwt_unpol_basic}
\end{equation}
using an integrated luminosity of 250 $fb^{-1}$ for each of $e^- \, \gamma$
and $e^+ \, \gamma$ modes.  As was shown in section
\ref{subsec:unpol_T-even}, imposing further cut of $p_{_T}(W) \geq 80 \; {\rm
GeV}$ and $\mid \sin{\phi_{HW}}\mid \; \geq \;0.4$ (the cut $\mathcal C_1$ of
Table \ref{rts_500_unpol}), enhances the relative effect of $\ibetw$. With
this set of cuts imposed, we get, instead
\begin{equation}
 \mid \mathcal I(\tbw) \mid \leq  0.092
\label{asymm_lim_rbwt_unpol_ptw_gt80_sin_gt.4}
\end{equation}
irrespective of the value of $\rbetw$ or $\Delta a_W$. 

We may also use the $e^+ \, \gamma$ mode to resolve between $\rbetw$ and
$\ibw$, which we have been unable to  working with  the $e^- \, \gamma$ mode
alone. As we have already seen (Equation \ref{asymm_unpol}), an asymmetry
constructed out of the azimuthal separation between the $H$ and the $W$ would,
in general, pick up the contributions proportional to $\rbetw$ and
$\ibw$. Read in conjunction with Equation \ref{asymm_Ac1_unpol}, we have,
for  a new asymmetry,
\begin{eqnarray}
 \mathcal A_{C2_{\rm unpol}} &=& \frac{ (\sigma_{++} -\sigma_{+- }) -
 (\sigma_{-+} -\sigma_{-- }) }{ (\sigma_{++} + \sigma_{+- }) +
 (\sigma_{-+} + \sigma_{-- })} \nonumber \\
&=& \frac{-4\; \left[\sigma_3 \;\mathcal I ( b_W)\right]_{e^-
 \gamma,\;\sin{ \phi_{HW}} > 0 } }{ 4\; \left[\sigma_0\; (1 + 2\;
 \Delta a_W) + \sigma_1\; \mathcal R(b_W) \right]_{e^- \gamma ,\;\sin{
 \phi_{HW}} > 0 } } \nonumber \\
&\simeq& - \ibw \, \left[\frac{\sigma_3}{\sigma_0} \right]_{e^- \gamma ,\;\sin{
 \phi_{HW}} > 0 }.,
\label{asymm_AC2} 
\end{eqnarray}
where
\begin{eqnarray}
\sigma_{++} &=& \sigma_{e^+\gamma, \; \sin{\phi_{HW} >0}} \nonumber \\
\sigma_{+-} &=& \sigma_{e^+\gamma, \; \sin{\phi_{HW} <0} }\nonumber \\
\sigma_{-+} &=& \sigma_{e^-\gamma, \; \sin{\phi_{HW} >0} }\nonumber \\
\sigma_{--} &=& \sigma_{e^-\gamma, \; \sin{\phi_{HW} <0} }
\label{def_asymm_unpol}
\end{eqnarray}
Use of this asymmetry  gives,  for the $3 \sigma$ bound,
\begin{equation}
 \mid \mathcal I(b_W) \mid \;\leq  \; 0.096
\label{lim_asymm_ibwt_unpol}
\end{equation}
for an integrated luminosity of 250 ${\rm fb}^{-1}$ per mode.

And finally to isolate $\rbetw$, we may construct yet another asymmetry 
\begin{eqnarray}
 \mathcal A_{C3_{\rm unpol }} &=& \frac{ (\sigma_{++} -\sigma_{-- }) -
 (\sigma_{+-} -\sigma_{-+ }) }{ (\sigma_{++} + \sigma_{-- }) +
 (\sigma_{+-} + \sigma_{-+ })} \nonumber \\
&=& \frac{4\; \left[\sigma_2 \;\mathcal R (\tbw)\right]_{e^+
 \gamma,\;\sin{ \phi_{HW}} > 0 }}{4\; \left[\sigma_0\; (1 + 2\;
 \Delta a_W) + \sigma_1\; \mathcal R(b_W) \right]_{e^- \gamma ,\;\sin{
 \phi_{HW}} > 0 } } \nonumber \\ 
&\simeq& - \rbetw \, \left[\frac{\sigma_4}{\sigma_0} \right]_{e^+ \gamma ,\;\sin{
 \phi_{HW}} > 0 }.
\label{asymm_AC3_unpol}
\end{eqnarray}
Thus for the $3 \sigma$  bound on $ \mid \mathcal R(\tbw) \mid $, we get
\begin{eqnarray}
 \mid \mathcal R(\tbw) \mid \; \leq \;  0.046
\label{asymm_lim_ibwt_unpol}
\end{eqnarray}
using an integrated luminosity of 250 ${\rm fb}^{-1}$ per mode.
\subsection{Polarized Beams}
\label{subsec:e+_pol}
It is easy to see that the introduction of non-zero beam polarization would
lead to a relation between $e^- \gamma$ and $e^+ \gamma$ cross-sections that is
analogous to that of Equation \ref{rel_e+e-_cross_sec_unpol}, namely
\begin{eqnarray}
\sigma_{e^+ \gamma,(P_L, P_b)} &= & \left[ (1+ 2\; \Delta a_W)\;
\sigma_0 +\mathcal R(b_W) \sigma_1+\mathcal R(\tbw)
\sigma_2\right]_{e^- \gamma\; ,(-P_L, -P_b)} \nonumber \\ 
& &-\left[\mathcal I(\tbw) \sigma_4 +\mathcal I( b_W) \sigma_3
\right]_{e^- \gamma\; ,(-P_L, -P_b)} 
\label{rel_pol_e+e-}
\end{eqnarray}
This is easy to understand since reversing both $P_L$ and $P_b$ results in
reversing the photon polarization while preserving the density
distribution (see Equation \ref{photon_spectrum}) We construct asymmetries to take
advantage of these. The best limits are obtained from
\begin{eqnarray}
 \mathcal A_{C1_{\rm pol}} &\equiv& \frac{ {\sigma_{e^- \gamma (+1, -0.8)}}
-{\sigma_{e^+ \gamma (-1, +0.8)} }  }{ {\sigma_{e^- \gamma (+1, -0.8)}}
+{\sigma_{e^+ \gamma (+1, -0.8) }} }\nonumber \\
&=& \frac{2\; \left[ \mathcal I (\tbw)\, \sigma_4 + \mathcal I(b_W)\,
 \sigma_3 \;\right]_{e^- \gamma, (+1, -0.8) }}{2\; \left[ (1 + 2\;
 \Delta a_W)\, \sigma_0 + \mathcal R(b_W)\, \sigma_1+ \mathcal R
 (\tbw) \sigma_2 \right]_{e^- \gamma, (+1, -0.8) }}  \nonumber \\
 &\simeq & \left[ \frac{\ibetw\; \sigma_2 + \ibw \; \sigma_3}{\sigma_0}
 \right]_{e^- \gamma, (+1, -0.8) }.
\label{asymm_AC1_pol}
\end{eqnarray} 
With only acceptance cuts imposed, $\ibw$ does not contribute to the
total cross-section and thus above asymmetry depends only on
$\ibetw$. We obtain for the $3 \sigma$ bound
\begin{equation}
 \mid \mathcal I(\tbw) \mid \;\leq \; 0.044
\label{lim_bwt_pol_e+e-}
\end{equation}
using integrated luminosity of $125\, {\rm fb}^{-1}$ per polarization
combination. This limit is stronger than the one obtained using unpolarized
photons (Equation \ref{asymm_lim_rbwt_unpol_basic}).

The other two couplings $\mathcal I(b_W)$ and $\mathcal R(\tbw)$ can
be also isolated and constrained using the mixed asymmetries as
before. To get limits on $\mathcal I(b_W)$, we construct the following
asymmetry
\begin{eqnarray}
 \mathcal A_{C2_{\rm pol}} &\equiv& \frac{ \left[ \sigma_{\sin{ \phi } > 0 }
-\sigma_{\sin{ \phi } < 0 } \right]_{e^+ \gamma, (+1, -0.8)}
-  \left[ \sigma_{\sin{ \phi } > 0} 
-\sigma_{\sin{ \phi } < 0 } \right]_{e^- \gamma, (-1, +0.8)} }{
 \left[ \sigma_{\sin{ \phi } > 0 }
+\sigma_{\sin{ \phi } < 0 } \right]_{e^+ \gamma, (+1, -0.8)}
+  \left[ \sigma_{\sin{ \phi } > 0} 
+\sigma_{\sin{ \phi } < 0 } \right]_{e^- \gamma, (-1, +0.8)} }
 \nonumber \\
&=& \left[\frac{4\;\mathcal I (b_W)\, \sigma_3 }{ 4\; \sigma_0\; (1 +
2\; \Delta a_W) + \sigma_1\; \mathcal R(b_W)} \right]_{e^+ \gamma
,\;\sin{ \phi } > 0 ,\;(+1,-0.8) } \nonumber \\ 
&\simeq& \ibw \; \left[\frac{ \sigma_3}{ \sigma_0}\right]_{e^+ \gamma ,\;\sin{
\phi } > 0 ,\;(+1,-0.8) }
\label{asymm_AC3_pol}
\end{eqnarray}
which gives  for the $3\sigma$ bound
\begin{equation}
 \mid \mathcal I( b_W) \mid \;\leq  \; 0.22
\label{asymm_lim_ibw_pol}
\end{equation}
using integrated luminosity of 125 ${\rm fb}^{-1}$ for each of the
polarization combinations.

To isolate $\rbetw$, we construct,
\begin{eqnarray}
 \mathcal A_{C3_{\rm pol}} &\equiv& \frac{ \left[ \sigma_{\sin{ \phi } > 0} 
-\sigma_{\sin{ \phi } < 0 } \right]_{e^+ \gamma,\; (+1, -0.8)}
+  \left[ \sigma_{\sin{ \phi } > 0 }
-\sigma_{\sin{ \phi } < 0 } \right]_{e^- \gamma,\; (-1, +0.8)}}{
 \left[ \sigma_{\sin{ \phi } > 0} 
+\sigma_{\sin{ \phi } < 0 } \right]_{e^+ \gamma,\; (+1, -0.8)}
+  \left[ \sigma_{\sin{ \phi } > 0} 
+\sigma_{\sin{ \phi } < 0 } \right]_{e^- \gamma,\; (-1, +0.8)} }
 \nonumber \\
&=& \left[ \frac{4\; \mathcal R (\tbw)\, \sigma_2  }{ 4\;
\sigma_0\; (1 + 2\; \Delta a_W) + \sigma_1\; \mathcal R(b_W)}
\right]_{e^+ \gamma ,\;\sin{ \phi } > 0 ,\;(+1,-0.8) } \nonumber \\
&\simeq& \rbetw \; \left[\frac{ \sigma_4}{ \sigma_0}\right]_{e^+ \gamma ,\;\sin{
\phi } > 0 ,\;(+1,-0.8) }
\label{asymm_AC2_pol}
\end{eqnarray}
which gives  for the $3\sigma$ bound
\begin{equation}
 \mid \mathcal R(\tbw) \mid \;\leq  \; 0.068
\label{asymm_lim_ibwt_pol}
\end{equation}
on using integrated luminosity of 125 ${\rm fb}^{-1}$ per polarization case.

 The comparison of limits on these couplings with the ones obtained using
 unpolarized photons is done in the Table \ref{lim_e+e-_asymm}. It may be
 noted that the polarized photons   help to improve the constraints on
 $\ibetw$ significantly over the unpolarized photons. For the other two
 couplings, however, the limits in fact worsen. This is due to reduction in
 the luminosity.
\begin{table}
\begin{center}
\begin{tabular}{|c|c|c|c|}
\hline Coupling & $ 3\; \delta \mathcal A$ bound & Observable Used &
Luminosity \\
& & &(in ${\rm fb}^{-1}$)\\
 \hline
$\mid \mathcal I(\tbw) \mid$ & 0.092 & $ \mathcal A_{C1_{\rm
unpol}}$ {\small with} $ \mathcal C_1$ & 250 \\
$\mid \mathcal I(\tbw) \mid$ & 0.044 & $ \mathcal A_{C1_{\rm
 pol}}$ {\small with acceptance cuts} & 125 \\
$\mid \mathcal I( b_W) \mid$ & 0.096 & $ \mathcal A_{C3_{\rm unpol}}$
 {\small with acceptance cuts} & 250 \\
$\mid \mathcal I(b_W) \mid$ & 0.220 & $ \mathcal A_{C3_{\rm pol}}$
 {\small with acceptance cuts} & 125 \\
$\mid \mathcal R( \tbw) \mid$ & 0.046 & $ \mathcal A_{C2_{\rm
 unpol}}$ {\small with acceptance cuts} & 250 \\
$\mid \mathcal R(\tbw) \mid$ & 0.068 & $ \mathcal A_{C2_{\rm
 pol}}$ {\small with acceptance cuts} & 125 \\
\hline
\end{tabular}
\end{center}
\caption{\em Comparison of limits on anomalous couplings at $3\sigma$
level using unpolarized and polarized beams.}
\label{lim_e+e-_asymm}
\end{table}
\section{Conclusions}
 Since the WWH couplings are not contaminated by the
ZZH couplings in the process studied hence $e\; \gamma$
colliders can be used to constrain the anomalous WWH couplings independent of
the ZZH couplings. Thus $e\; \gamma$ colliders are better equipped than $e^+
\; e^-$ colliders to study these couplings. 

Comparing our results to those of Ref. \cite{bcgs}, we find that we obtain
better individual limits for all couplings, bar $\Delta a_W$, using
unpolarized photon beams. These limits on $\rbw, \ibetw$ and $\rbetw$ become
stronger with the use of polarized photons. Polarized photons can be also used
to derive constraints on the $\ibw-\rbetw$ plane, something not possible with
the use of unpolarized photons. Furthermore, in Ref. \cite{bcgs}, the authors
were unable to construct observables that depend on only one of the
couplings. Hence their limits on WWH couplings are not independent of each
other. However, using the process $e^- \;\gamma \lra \nu \; W^- \; H$ in
conjunction with the conjugate process $e^+ \; \gamma \lra \bar \nu \; W^+ \;
H$, and using the $P \; \hat T$ properties of various contributions to the
total rate, we are able to construct observables that are function of only one
of the couplings.  Thus we are able to derive constraints on each of the
couplings $\ibetw$, $\ibw$ and $\rbetw$ independent of the value of any other
coupling. $\Delta a_W$ and $\rbw$, however, cannot be constrained independent
of each other.

We also conclude that once both  $e^-\; \gamma$ and $e^+\; \gamma$ initial
states can be used, beam polarization  does not give any significant advantage
and strong limits may be obtained with the use of unpolarized photons alone.
%
\section*{Acknowledgment}
 One of us (Mamta) wants to thank Department of Science and
Technology, New Delhi for providing partial support during the 
course of this work under the grant no. SP/S2/K-20/99.

%
\end{document}